\documentclass[preprint,prl,showpacs,superscriptaddress,
draft,amsmath,amssymb]{revtex4}

\usepackage{graphicx}% Include figure files
\usepackage{dcolumn}% Align table columns on decimal point
\usepackage{bm}% bold math
\usepackage{color}

\makeatletter
\def\gsim{\compoundrel>\over\sim}

\def\compoundrel#1\over#2{\mathpalette\compoundreL{{#1}\over{#2}}}
\def\compoundreL#1#2{\compoundREL#1#2}
\def\compoundREL#1#2\over#3{\mathrel
    {\vcenter{\hbox{$\m@th\buildrel{#1#2}\over{#1#3}$}}}}
\makeatother

\begin{document}

\title{Femtosecond core level photoemision spectroscopy on 
$1T$-TaS$_2$ using 60 eV laser source}

\author{K. \surname{Ishizaka} }
\altaffiliation{Present address: Department of Applied Physics, University of Tokyo, Tokyo 113-8656, Japan}
\affiliation{Institute for Solid State Physics, University of Tokyo,
 Kashiwa, Chiba 277-8581, Japan}
\affiliation{CREST, JST, Tokyo 102-0075, Japan}

\author{T. \surname{Kiss}}
\altaffiliation{Present address: Graduate School of Engineering Science, Osaka University, Osaka 560-8531, Japan}
\affiliation{Institute for Solid State Physics, University of Tokyo,
 Kashiwa, Chiba 277-8581, Japan}
\affiliation{CREST, JST, Tokyo 102-0075, Japan}

\author{T. \surname{Yamamoto}}
\affiliation{Institute for Solid State Physics, University of Tokyo,
 Kashiwa, Chiba 277-8581, Japan}
\affiliation{Department of Applied Physics, Tokyo University of Science, Shinjuku-ku, Tokyo 162-8601, Japan}

\author{Y. \surname{Ishida}}
\affiliation{Institute for Solid State Physics, University of Tokyo,
 Kashiwa, Chiba 277-8581, Japan}
\affiliation{CREST, JST, Tokyo 102-0075, Japan}

\author{T. \surname{Saitoh}}
\affiliation{Department of Applied Physics, Tokyo University of Science, Shinjuku-ku, Tokyo 162-8601, Japan}

\author{M. \surname{Matsunami}}
\altaffiliation{Present address: UVSOR Facility, Institute for Molecular Science, Okazaki 444-8585, Japan}
\affiliation{Institute for Solid State Physics, University of Tokyo,
 Kashiwa, Chiba 277-8581, Japan}
\affiliation{RIKEN SPring-8, Sayo-gun, Hyogo 679-5148, Japan}

\author{R. \surname{Eguchi}}
\altaffiliation{Present address: Research Laboratory for Surface Science, Okayama University, Okayama 700-8530, Japan}
\affiliation{Institute for Solid State Physics, University of Tokyo,
 Kashiwa, Chiba 277-8581, Japan}
\affiliation{RIKEN SPring-8, Sayo-gun, Hyogo 679-5148, Japan}

\author{T. \surname{Ohtsuki}}
\affiliation{RIKEN SPring-8, Sayo-gun, Hyogo 679-5148, Japan}

\author{A. Kosuge}
\altaffiliation{Present address: Advanced Photon Research Center, Japan Atomic Energy Agency, Kyoto 619-0215, Japan}
\affiliation{Institute for Solid State Physics, University of Tokyo,
 Kashiwa, Chiba 277-8581, Japan}

\author{T. Kanai}
\affiliation{Institute for Solid State Physics, University of Tokyo,
 Kashiwa, Chiba 277-8581, Japan}

\author{M. Nohara}
\affiliation{Department of Physics, Okayama University,
Kita-ku, Okayama 700-8530, Japan}

\author{H. Takagi}
\affiliation{School of Frontier Sciences, University of Tokyo,
 Kashiwa, Chiba 277-8581, Japan}

\author{S. Watanabe}
\altaffiliation{Present address: Research Institute for Science and Technology, Tokyo University of Science, Chiba 278-8510, Japan}
\affiliation{Institute for Solid State Physics, University of Tokyo,
 Kashiwa, Chiba 277-8581, Japan}

\author{S. \surname{Shin}}
\affiliation{Institute for Solid State Physics, University of Tokyo,
 Kashiwa, Chiba 277-8581, Japan}
\affiliation{CREST, JST, Tokyo 102-0075, Japan}
\affiliation{RIKEN SPring-8, Sayo-gun, Hyogo 679-5148, Japan}

\begin{abstract}

Time-resolved photoelectron spectroscopy (trPES) can directly detect transient electronic structure, thus bringing out its promising potential to clarify nonequilibrium processes arising in condensed matters. Here we report the result of core-level (CL) trPES on 1$T$-TaS$_2$, realized by developing a high-intensity 60 eV laser obtained by high-order harmonic (HH) generation. Ta$4f$ CL-trPES offers the transient amplitude of the charge-density-wave (CDW), via the site-selective and real-time observation of Ta electrons. The present result indicates an ultrafast photoinduced melting and recovery of CDW amplitude, followed by a peculiar long-life oscillation (i.e. collective amplitudon excitation) accompanying the transfer of 0.01 electrons among adjacent Ta atoms. CL-trPES offers a broad range of opportunities for investigating the ultrafast atom-specific electron dynamics in photo-related phenomena of interest.

\end{abstract}

\maketitle

%\draft
%\narrowtext

%%%%%%%%%%%%%%%%%%%%%
%%Introduction%%%%%%%%
%%%%%%%%%%%%%%%%%%%%%%

%%%%%%%%%%%%%%%%%%%%%
%%EXPERIMENTAL%%%%%%%%
%%%%%%%%%%%%%%%%%%%%%%

%%%%%%%%%%%%%%%%%%%%%%%%%%%%%%%%%
%%%%%Results Figure 1%%%%%%%%%%%%%
%%%%%%%%%%%%%%%%%%%%%%%%%%%%%%%%%%

Photoelectron spectroscopy (PES) 
is a versatile probe for directly studying 
the electronic structure of matters.
In particular, 
core-level (CL) PES using x-rays as the light source 
can probe site-specific electronic states of 
individual atoms in compounds, reflecting the local chemical environment. 
%which is  
%relevant to a wide variety of phenomena arising in fields of  
%chemistry, physics, material science, and biology. 
Integration of pump-probe method into CL-PES 
further opens the new frontier, the real-time atom-specific investigation of ultrafast dynamics
in complex phenomena such as photochemical and photobiological reactions, and 
photoinduced phase transitions \cite{zewail00,dwayne08,PIPT3}. 
Here we study the ultrafast real-time photo-response of the peculiar
CDW Mott-insulator state in a layered dichalcogenide material $1T$-TaS$_2$ {\it via} the Ta $4f$ CL-PES, which directly gives us quantitative and site-selective information on the excitation/relaxation processes of the CDW.

1$T$-TaS$_2$ shows a complex sequence of temperature ($T$) -dependent
CDW transitions. CDW is already formed at 300 K, with its amplitude nearly commensurate (NC-CDW) to the triangular lattice. At lower $T$, it experiences the NC-CDW to commensurate CDW (C-CDW)  transition at around 190 K,  with a 1-order magnitude of jump in resistivity. This is explained as the Mott transition due to the narrowing of Ta 5$d$ valence band at the NC-CDW to C-CDW transition \cite{fazekas79}. This C-CDW Mott insulator phase is known to be fragile against external perturbations, such as electron beam irradiation \cite{mutka81} or contamination \cite{zwick98}, and easily turns into a disordered metallic state. Also by applying pressure, the C-CDW phase quickly loses its long-range coherence and shows the superconducting phase transition \cite{sipos08}. Thus, 1$T$-TaS$_2$ possesses a complex electronic state where rather strong electron-phonon and electron-electron couplings are inherent and entangled. 
Ta4$f$ CL spectrum directly reflects the nonequivalent Ta sites introduced by the CDW modulation, since its binding energy ($E_B$) is very sensitive to the local charge density of each Ta \cite{pollak81}.
Thus, Ta4$f$ time-resolved PES (trPES) can directly track the Ta-site-specific transient photoinduced response of the peculiar CDW state.

Recently reported UV-laser based trPES method successfully detects photoinduced transient electronic structures \cite{perfetti06,schmitt08}, nevertheless, 
thus far have been limited in a very narrow energy region at the Fermi level due to the lack of high enough photon energy. 
Also considering that the pulse duration of the synchrotron x-ray is limited in the picosecond range, the progress of novel high-intensity short-pulsed x-ray sources is necessary. 
Very recently developed x-ray free-electron-laser (xFEL) is one such strong 
candidate for CL-trPES light source \cite{pietzsch08,hellmann10}, 
though presently there is a 
difficulty of space-charge effect due to its rather low 
repetition rate ($\sim $Hz) and also the effective 
time resolution affected by the 
temporal jitter of electron pulses.  
Another potential is the coherent tabletop x-ray source based on the high-harmonic (HH) generation of a short-pulsed optical laser \cite{avila08}.

Here, we have constructed a new PES system utilizing the 
HH generation of high-repetition-rate (1 kHz) 
Ti-Sapphire laser, as shown in Fig. 1.
We use the second harmonic (SH) with a photon energy $\omega= 3.15$ eV (394 nm), generated by a 0.5-mm thick $\beta $ - BaB$_2$O$_4$ (BBO) crystal, as a driving source for HH generation. 
 The average power, repetition rate, and the pulse duration of the SH light 
are 1.5 W,  1 kHz, and 170 fs, respectively. 
We use SH rather than the fundamental to obtain stronger harmonics ($\sim$nJ/pulse on the target) due to increased atomic dipole moment by short wavelength drive and to double the harmonic separation (6.3 eV)  which reduces the disturbance by neighboring harmonics \cite{sekikawa04}.
The beam is focused at a pulsed Ne gas jet (pulse width $\sim$0.2 msec) synchronized with the repetition of the laser. Here we use the 19th harmonic, 19$\omega = 60$ eV, selected by a pair of wave-selective multilayer (Mo/Si) mirrors with other HHs suppressed. 
A hemispherical electron analyzer is used for photoelectron spectroscopy.  
The total energy resolution of this PES system estimated from the observed Fermi cutoff of Au spectrum (Fig. 1 inset) is 230 meV.

Surrounded by S octahedrons, Ta atoms in 1$T$-TaS$_2$ form a two-dimensional triangular network as shown in Fig. 2(a).
The $T$-dependent Ta4$f_{7/2}$ spectrum obtained using synchrotron light source ($h\nu = 1200$ eV, energy resolution 200 meV) is shown in Fig. 2(b).  At 373 K where the CDW is incommensurate, the Ta4$f$ spectrum shows a broad peak feature. At 30 K C-CDW state, it clearly splits into two sharp peaks reflecting the $\sqrt{13} \times \sqrt{13}$ structure, where the 13 Ta atoms form a star-of-David cluster \cite{wilson78} as shown in Fig. 2(a). 
In this scheme, there are 3 nonequivalent Ta sites; the one at the center (site $a$) and 2 respective sets of 6 atoms forming inner- ($b$) and outer- ($c$) rings. According to a model calculation explicitly taking into account of this C-CDW modulation \cite{smith85}, the local integrated densities at sites $a$, $b$, and $c$ are estimated to be 1.455, 1.311, and 0.611 electrons ($e$), 
respectively.
Thus, the 2 peaks at higher and lower $E_B$ in observed Ta4$f$ CL spectra correspond to sites $c$ and $b$, whereas the peak for site $a$ is known to appear at a lower $E_B$ with very small intensity \cite{hughes95}. 
\textcolor{black}{Here the split energy between these 2 peaks is linearly related to the amount of electron transfer between sites $c$ and $b$ ({\it i.e.} 0.7$e$ according to the abovementioned calculation \cite{smith85}), thus semiquantitatively reflecting the amplitude of the CDW.}    
The $T$-dependent Ta4$f_{7/2}$ CL spectrum obtained by the new 60 eV laser-HH PES apparatus is shown in Fig. 2(c). The energy resolution was set to 470 meV to obtain enough S/N ratio. It is also very similar to that using synchrotron light source, when we take into account of the difference of the energy resolutions.

Figure 3(a) shows the trPES result of Ta $4f$ CL for 1$T$-TaS$_2$, obtained at 30 K. The pump intensity used here is 2 mW, corresponding to pulse density of 1 mJ/cm$^{2}$. The delay time ($t$) represents the difference of the probe and pump pulse arrival at the sample surface. The time origin $t=0$ is defined as the point where the photoinduced spectral change takes its maximum. At $t = -0.5$ ps, the spectrum clearly shows 2 peak structures at around  $E_B \sim 23.9$ eV (site $c$) and 23.2 eV (site $b$). Its spectral shape coincides well with that obtained without pumping, indicating that the heating effect while pumping does not seriously affect this measurement. On approaching $t=0$, peak $c$ especially shows a clear change, starting to collapse and get closer to peak $b$. Such an excited electronic state already disappears at $t= 1$ ps, showing rather fast relaxation of this photoinduced phenomenon (See Supplementary for a movie file)
To quantitatively analyze the data, we fit the CL spetra using 4 Voigt functions (2 peaks at higher binding energy corresponding to the Ta $4f_{5/2}$ levels, not shown). Voigt function represents the convolution of Gaussian to Lorenzian peaks, where the Gaussian widths are fixed to the energy resolution of 470 meV. For simplicity, the Lorenzian widths ($\Gamma$) corresponding to the line widths of respective CL are set to take a common value for peak $c$ and $b$, and the asymmetry factor is fixed to zero.
Red curves in Fig. 3(a) are the fitting to respective PES spectra, whereas blue and green markers show the peak positions obtained by the fitting.

Since the split energy ($\Delta $) between peaks $b$ and $c$ should be linearly related to the amplitude of C-CDW, we plot in Fig. 3(c) the subtraction of the obtained peak positions as $\Delta (t)$, together with the spectral line width $\Gamma (t)$ in Fig. 3(d). At $t\sim 0$, $\Delta$ abruptly decreases from 0.745 to 0.667 eV \textcolor{black}{indicating the suppression of CDW amplitude by 10\%}, and quickly recovers within 1 ps.  Regarding the line width $\Gamma$, it increases from 250 to 360 meV at $t\sim 0$, and also seems to recover in $\sim 1$ ps within error bars. 
Here, the transient curve of $\Delta (t>0)$ can be described by two major processes, a rather fast relaxation ($t< 1$ ps) and the oscillatory component surviving well beyond 1 ps after pumping. Following the analysis performed in a number of CDW systems \cite{demsar02}, here we express the relaxation process of $\Delta (t)$ as        
$$\Delta (t) = \Delta_0 - \delta\Delta_e e^{-t/\tau_{e}} -\delta\Delta_p e^{-t/\tau _{p}}\cos(2\pi f_0 t + \phi)  $$
with $\Delta_0= 0.745$ eV, $\delta\Delta_e = 0.08\pm 0.01$ eV, $\delta \Delta_p = 0.020\pm 0.005$ eV, $f_0 =2.44\pm 0.06$ THz, $\tau_e = 0.46\pm 0.05 $ ps, and $\tau_p \gsim 3$ ps. $\Delta _0 $ is the initial split energy, whereas the second and third terms correspond to the ultrafast relaxation and oscillatory component, respectively. The obtained function is depicted in Fig. 3(c) as the red curve, which well agrees at least with the data taken for the temporal region of 1.5 ps after pumping. The frequency of the oscillatory component coincides with the $A_{1g}$ CDW amplitude mode appearing at 81 cm$^{-1}$ ($=2.45$ THz) in Raman scattering \cite{sugai85} and infrared \cite{uchida81} spectra. Thus, the  $\delta \Delta _p$ term shows that the CDW amplitude oscillates in real time after pumping, corresponding to the CDW amplitudon excitation. 
The magnitude of this oscillation can be estimated by $\delta\Delta_p / \Delta_0$, {\it i.e.} 2.7\% of the initial state C-CDW amplitude.

We can further discuss the CDW amplitude semiquantitatively in terms of the relative difference of electron numbers between sites $b$ and $c$, $n_b-n_c$, by assuming the electronic configuration given by the model calculation \cite{smith85}. According to this model, $n_b-n_c$ is given by 0.7$e$ at the initial state. Thus, we can estimate the $t$-dependent variation of $n_b-n_c$ by $\delta (n_b-n_c) = \frac{\Delta(t) - \Delta_0}{\Delta_0}\times 0.7e $ [see the right axis in Fig. 3(c)].
From the above analysis, we can conclude that the photoinduced suppression of the CDW amplitude \textcolor{black}{(by $\sim 0.07e$)} recovers within a relaxation time scale of $\sim \tau_e =0.46$ ps, while the excitation of amplitudon involving the oscillatory electron transfer of $\sim 0.01e$ lasts for at least 3 ps.
It is interesting to note that the time scale of $\tau_e$ is fairly close to the oscillation of amplitudon ($f_0 ^{-1} =0.41$ ps), which may be indicating the possibility that the energy of the thermalized electrons is relaxed through the excitation of the amplitudons. 
This transient behavior agrees well with that observed by pump-probe measurements of reflectivity \cite{demsar02} and near-$E_F$ trPES \cite{perfetti06,perfetti08}, showing the big sub-ps response at $t \sim 0$ together with the temporal oscillation existing for more than 20 ps.
Particularly, the near-$E_F$ trPES reports the simultaneous suppression of lower Hubbard band and the increase of the intensity at $E_F$ \cite{perfetti06,perfetti08} at $0<t< 0.68$ ps, which they attribute to the photoinduced insulator-metal transition. 
The present result thus indicates that such photoinduced metal transition indeed accompanies the suppression of CDW amplitude.

The trPES CL spectrum includes the transient site-selective information of Ta in CDW photo-melting and recovery process. In Fig. 4(a), the Ta $4f$ CL spectra before ($t=-0.5$) and immediately after pumping ($0$ ps) obtained by two different pumping powers are shown. The $T$-dependent CL spectrum measured using $h\nu = 1200$ eV is also shown in Fig. 4(b). For the $T$-dependent measurement, where the thermal equilibrium state is maintained while CDW melting, the CL spectrum shows that both peaks $b$ and $c$ get broader and closer to each other on increasing $T$ above 200 K, indicating the gradual suppression of C-CDW clusters and the introduction of domain boundaries in NC-CDW state \cite{yamamoto83}. On the other hand, the photoinduced spectral change mainly appears as the quick collapse and red-shift of peak $c$ while site $b$ remains mostly unchanged, as observed in both independent sets of data. It shows that the photoexcitation almost selectively modifies the electronic state and increases the number of electrons at site $c$. It is worth noting that the amplitudon oscillation also mainly shows up in the peak position of site $c$ [Fig. 3(b), see also Supplementary], indicating that they are strongly coupled. 

To naively account for this phenomenon, first we should note that the optical pumping with 3.15 eV light is mostly characterized by charge transfer excitation from S$3p$ to Ta$5d$, according to the band calculation \cite{mattheis73}. Besides, it is reported that the Ta$5d$ band gets reconstructed into 3 subbands due to the C-CDW formation \cite{zwick98} as shown schematically in Fig. 4(c), where the occupied lower 2 bands are expected to mainly consist of Ta sites $a$ and $b$, respectively \cite{smith85}. Since site $c$ with the highest binding energy is the most unoccupied, the pumping should induce the electron transfer from S$3p$ to Ta site $c$ as compared with sites $a$ and $b$. This may be the origin of the site selective photoresponse as observed. 
We would like to also stress that site $c$ belongs to the outer-ring of the Star-of-David in C-CDW state, where the Ta position experiences the largest deformation (up to 7\%) by the CDW transition \cite{brouwer80}. Naturally, the amplitudon oscillation is strongly coupled to this deformation. This situation may be the origin of the photoinduced amplitudon oscillation mainly arising in the binding energy for site $c$.  
Future trPES experiments with more developed higher-energy HH laser or xFEL will further enable us to directly observe S$2p$ CL ($E_B \sim$ 160 eV), which should lead to the deeper understanding of this photoinduced phenomenon, including the transient electron transfers among respective Ta and S sites.

In summary, 
we constructed a pump-probe trPES system with a tabletop coherent femtosecond x-ray laser, and investigated the transient photoresponse of characteristic CDW Mott insulator 1$T$-TaS$_2$. The present Ta$4f$ trPES result successfully offers the Ta-site-specific information on the photoinduced ultrafast melting/recovery of CDW amplitude and the collective amplitudon excitation. Microscopic electronic interpretation of such collective CDW excitations has become possible for the first time, owing to the site-selective and real-time observation of Ta electrons.

\newpage

\begin{figure}[htbp!]
\caption{
(a) Schematic diagram of the trPES apparatus.  
(b) Photoemission spectrum from Au obtained at 300 K using this HH-laser PES system.  Inset shows the Fermi cutoff fitted by gaussian-convolved Fermi-Dirac function. By fitting, the total energy resolution is estimated to be 230 meV.   
\label{fig1}
}
\end{figure}

\begin{figure}[htbp!]
\caption{
(a) Schematic picture of Ta atom network in the high $T$ phase (upper) and low $T$ commensurate CDW state (lower).
(b) $T$-dependent Ta$4f_{7/2}$ PES spectra obtained by synchrotron x-ray light source ($h\nu = 1200$ eV, SPring-8 BL17SU).
(c) Ta$4f_{7/2}$ PES spectra recorded at 300 and 190 K, using HH 60 eV laser PES system. 
\label{fig2}
}
\end{figure}

\begin{figure}[htbp!]
\caption{
(a) Transient PES spectra for temporal region of $-0.5 \leq t \leq 1.6$. 
Red curves are the fitting to respective PES spectra, blue and green markers show the peak positions obtained by the fitting.
(b) PES intensity image as a function of binding energy and time delay clearly shows the shift and smearing of peak c at around $t=0$.   
(c) Split energy between the doublet peaks plotted as a function of delay time. Gray and red curves indicate a guide for eyes and the calculated function, respectively.  (d) Line width of the doublet peaks as a function of delay time.
\label{fig3}
}
\end{figure}

\begin{figure}[htbp!]
\caption{
(a) trPES spectra recorded before ($t=-0.5$ ps) and after ($t=0$) pumping, obtained by two sets of measurements with $I_{\rm pump}$ = 1 and 1.5 mJcm$^{-2}$, respectively. (b) $T$-dependent PES spectra obtained by $h\nu = 1200$ eV, broadened by gaussian convolution to match the energy resolution of trPES data.
(c) Schematic of the Ta$5d$ and S$3p$ valence bands and the optical transition by pumping (orange arrow).  The top subband forms a Mott gap at $E_F$ below the Mott transition temperature. 
(d)  Schematic of a star-of-David cluster.
\label{fig4}
}
\end{figure}

\end{document}